\documentclass[12pt,preprint]{aastex}

\usepackage{epsf}
\newcommand{\icar}{{\it Icarus, }}
\newcommand{\mnr}{{\it MNRAS, }}
\newcommand{\sci}{{\it Sci, }}
\newcommand{\ana}{{\it A\&A, }}

\begin{document}

\title{Migration and the formation of systems of hot super--Earths and
Neptunes}

\author{Caroline Terquem\altaffilmark{1,2}} 

\affil{Institut d'Astrophysique de Paris, UMR7095 CNRS,
Universit\'e Pierre \& Marie Curie--Paris~6, 98bis boulevard Arago,
75014 Paris, France}
\email{terquem@iap.fr}

\and

\author{John C. B. Papaloizou} 

\affil{Department of Applied Mathematics and Theoretical Physics,
University of Cambridge, Centre for Mathematical Sciences, Wilberforce
Road, Cambridge, CB3 0WA, UK}
\email{J.C.B.Papaloizou@damtp.cam.ac.uk}

\altaffiltext{1}{Universit\'e Denis Diderot--Paris 7, 2 Place Jussieu, 75251
Paris Cedex 5, France}

\altaffiltext{2}{Institut Universitaire de France}

%\begin{document}

%\maketitle

\begin{abstract}
The existence of extrasolar planets with short orbital periods
suggests that planetary migration induced by tidal interaction with
the protoplanetary disk is important. Cores and terrestrial planets
may undergo migration as they form. In this paper we investigate the
evolution of a population of cores with initial masses in the range
0.1--1~earth mass embedded in a disk.  Mutual interactions lead to
orbit crossing and mergers, so that the cores grow during their
evolution.  Interaction with the disk leads to orbital migration,
which results in the cores capturing each other in mean motion
resonances.  As the cores migrate inside the disk inner edge,
scatterings and mergers of planets on unstable orbits together with
orbital circularization causes strict commensurability to be lost.
Near commensurability however is usually maintained.  All the
simulations end with a population of typically between two and five
planets, with masses depending on the initial mass.  These results
indicate that if hot super--Earths or Neptunes form by mergers of
inwardly migrating cores, then such planets are most likely not
isolated.  We would expect to always find at least one, more likely a
few, companions on close and often near--commensurable orbits.  To
test this hypothesis, it would be of interest to look for planets of a
few to about 10~earth masses in systems where hot super--Earths or
Neptunes have already been found.
\end{abstract}


\section{Introduction} 
\label{intro}

The recent announcements of the detection of extrasolar planets of a
few earth masses (M$_\oplus$) (OGLE--05--390L~b, 5.4~M$_\oplus$,
Beaulieu et al. 2006; Gliese~876~d, 7.3~M$_\oplus$, Rivera et
al. 2005) gives support to the initial solid core accumulation model
for planet formation.  Although it has been proposed that such
``super--Earths'' could have formed through gravitational instability
in a gaseous protoplanetary disk leading to a giant planet which
subsequently lost its gaseous envelope through the action of external
UV radiation (Boss 2006), a formation mechanism through the
accumulation of planetesimals seems more natural.  In addition to
these super--Earths, nine planets with a mass comparable to that of
Uranus or Neptune (in the range 10.5 to 18.5~M$_\oplus$) have been
reported.

OGLE--05--390L~b is at a distance of 2.1~astronomical unit (au) from
the central star and was detected through microlensing.  Gliese~876~d,
detected through radial velocity measurements, is at 0.02~au from its
parent star.  Among the nine  planets with masses similar to that
of Neptune, four are within
0.1~au from the central star.

Gliese~876 is a M--type star.  The temperature at 0.02~au from the
star is therefore low enough for heavy elements to condense.  Thus it
is possible to consider that Gliese~876~d formed {\em in situ} by
accumulation of heavy material that spiraled in with the gas through
the circumstellar disk. However, the existence of close orbiting giant
planets, the so--called 'hot Jupiters' and 'hot Neptunes', has been
taken as an indication of the operation of large scale migration
induced by the interaction of recently formed protoplanets with
protoplanetary disks (e.g., Papaloizou~\& Terquem~2006 and references
therein). Such migration processes may have operated during the
formation of lower mass planets as well.  In particular, planets in
the earth mass range are expected to undergo type~I migration (e.g.,
Ward~1997).  Accordingly, here we envisage a scenario in which cores
assemble further away from the central star, and migrate inwards due
to tidal interaction with the disk.  The disk is supposed to be
truncated at some inner edge so that the planets do not fall onto the
star.  Mutual perturbations of the cores leads to orbit crossing,
collisions and possibly mergers during the migration phase, resulting
in the formation of a smaller number of more massive planets on short
period orbits.

Such a scenario has been considered by Brunini \& Cionco (2005), who
calculated by means of $N$--body simulations the evolution of
100~protoplanets of 0.5~M$_\oplus$ together with 200~planetesimals of
0.1~M$_\oplus$ subject to their mutual interaction and the tidal
interaction with the disk.  Focusing on the mass and semi--major axis
of the final largest solid core, they concluded that Neptune--like
planets on short orbits should be common.

Here we focus on the final configuration of the multi--planet systems
we obtain.  Evolution of an ensemble of cores in a disk almost always
leads to a system of a few planets, with masses that depend on the
total initial mass, on short orbits with mean motions that frequently
exhibit near commensurabilities and, for long enough tidal
circularization times, apsidal lines that are locked together.
Starting with a population of 10 to 25 planets of 0.1 or 1~M$_\oplus$,
we end up with typically between two and five planets with masses of a
few tenths of an earth mass or a few earth masses, depending on the
total initial mass, inside the disk inner edge.  Interaction with the
central star leads to tidal circularization of the orbits which,
together with possible close scatterings and final mergers, tends to
disrupt mean--motion resonances that are established during the
migration phase.  The system, however, often remains in a
configuration in which the orbital periods are close to
commensurability.  Apsidal locking of the orbits, if established
during migration, is often maintained through the action of these
processes.

The plan of the paper is as follows.  In section~\ref{sec:secular}, we
review studies of the evolution of migrating resonant planets.  We
also show that tidal circularization by the central star results in
the disruption of strict commensurability, although the mean motions
may remain nearly commensurable and apsidal lines remain locked.  In
section~\ref{sec:model}, we describe the model we use to follow the
evolution of a population of planets, and describe our initial
conditions.  To calculate the mutual interactions between the cores,
we use a $N$--body code.  A dissipative force is included to model the
tidal interaction with the disk, which leads to orbital decay and
eccentricity and inclination damping.  Relativistic effects and tidal
interaction with the central star are also included, as they affect
the eccentricity of the orbits close to the star.  We also incorporate
the possibility of corotation torques acting in the edge region. The
potential importance of these in reversing type~I migration has been
indicated by Masset et al.~(2006). We discuss the effects of such
torques on eccentric orbits.  In section~\ref{sec:results}, we
describe our results.  As mentioned above, all the runs end with a few
planets on close orbits inside the inner cavity that frequently
exhibit near commensurability and apsidal line locking.  We study the
effect of varying the circularization timescale and of a hypothetical
reversal of the torque near the disk inner edge due to corotation or
other effects.  Finally, in section~\ref{sec:disc}, we summarize and
discuss our results.

\section{Migration of planets and orbital resonance}

\label{sec:secular}

\subsection{Resonant capture during migration}

The existence of commensurabilities among the mean motions of pairs of
satellites of Jupiter and Saturn is believed to be the result of
capture into resonances following the differential expansion of their
orbits induced by the dissipation of the tides raised in the central
planet (Greenberg et al. 1972, Greenberg 1973).  Once established,
such commensurabilities are stable due to the secular transfer between
the satellites of the angular momentum fed into the satellite system
by the tides (Goldreich 1965).  Planets in a disk may also get locked
into a resonance if their semi--major axes evolve at a different rate
causing their orbits to approach each other.  Melita \& Woolfson
(1996; see also Haghighipour 1999) were first to consider such a
scenario to explain near--commensurability between the periods of the
major pla\-nets of the solar system.  In their study, the evolution of
the semi--major axes was assumed to be caused by accretion of gas by
the planets and dynamical friction.  Kley (2000) subsequently studied
the evolution of two Jupiter--like planets embedded in a
protoplanetary disk that underwent orbital migration due to tidal
interaction with the disk.  Formation and maintenance of
commensurabilities in a system of migrating planets in this type of
simulation was subsequently reported by Masset \& Snellgrove (2001)
and Snellgrove et al. (2001).  Since then, different studies,
motivated by the observation of extrasolar planetary systems
exhibiting commensurabilities, have shown that capture of giant
planets into resonances during migration is a natural expectation
(Nelson \& Papaloizou 2002, Lee \& Peale 2002, Kley et al. 2004).  The
planets subsequently migrate maintaining the commensurability.  The
capture into resonance of migrating planets in the earth mass range
has also been studied (Papaloizou \& Szuszkiewicz 2005, McNeil et
al. 2005, Cresswell \& Nelson 2006).

Once an embedded pair of planets is in resonance, the resonant angles
(see below) and the angular difference of the apsidal lines are
generally found to librate about fixed values.  Note that the apsidal
lines need not necessarily be aligned or anti--aligned.  The relative
orientation of the orbits may be phase--locked at an angle that
differs from 0 or 180~$\deg$, depending on their eccentricities and
the mass of the planets (Beaug\'e et al. 2003, Kley et al. 2004).

In this paper, we study the commensurabilities that are established
when a population of protoplanets with masses on the order of an earth
mass migrate together through a protoplanetary disk.  As they migrate,
the planets undergo collisions which are assumed to result in
mergers. This causes their masses to grow with time.  We suppose that
the protoplanetary disk has an inner edge interior to which is a
cavity inside which disk--planet interactions and induced orbital
migration cease. Observations suggest the existence of magnetospheric
cavities of this type with their extent controlled by the magnetic
field of the central star (e.g., Bouvier et al. 2006). Inner disk
boundaries in the range 0.05--0.1~au might be expec\-ted.  Once the
planets enter such a cavity, both additional collisions and mergers as
well as tidal interaction with the star cause strict commensurability
to be lost, although near--commensurability and locking of the apsidal
lines of the orbits of pairs of planets may be maintained under the
action of both these processes.  Similar effects resulting from tidal
interaction have been noted in the context of the solar system (e.g.,
Dermott et al. 1988) and in the context of extrasolar systems of giant
planets (Novak et al. 2003).  We now consider the dynamics of two
planets in mean motion resonance, with locked apsidal lines, that are
subject to tidal interaction with the central star that causes orbital
circularization and show how strict commensurability is lost while the
relative orientation of the apsidal lines can be maintained.

\subsection{Loss of commensurability through tidal dissipation}

\label{sec:disruption}

Once a planet has migrated to small enough radii, tidal interaction
with the central star becomes significant.  If we assume that the
rotation period of the central star is longer than the orbital period
of the planet, which is expected for planets with orbital periods of
$\sim$~4 days or less, tidal interaction between the star and the
planet leads to eccentricity damping and orbital decay.  The
consequent reduction in semi--major axis would tend to cause any
previously formed commensurability to be lost.  Here we neglect the
tides raised on the star by the planet, as we only consider planets of
a few earth masses for which such tides are not expected to be
significant (Goldreich \& Soter 1966). To get some insight into the
dynamics of the system, we consider a simple model in which there are
only two planets orbiting a star of mass $M_\star .$ We denote by
$m_i$, $a_i$, $e_i$ and $n_i$ the mass, semi--major axis, eccentricity
and mean motion of the inner planet $(i=1)$ and the outer planet
$(i=2).$ The two planets are presumed to be in a mean motion resonance
so that $n_1/n_2=(p+q)/p$, where $p$ and $q$ are two integers.  To
simplify the discussion, we consider only the $e_i^q$--eccentricity
resonances with $q=1.$ The associated resonant angles are $\Phi_1=p
\lambda_1 -(p+1) \lambda_2 + \tilde{\omega}_1$ and $\Phi_2=p \lambda_1
- (p+1) \lambda_2 + \tilde{\omega}_2$ where, for $(i=1,2),$
$\lambda_i$ are the mean longitudes and $\tilde{\omega}_i$ are the
arguments of pericentre.

The rates of change of the semi--major axes and eccentricities for
planets $i= (1,2)$ induced by the resonant interaction can be found
from the following equations (e.g., Dermott~et al.~1988):
\begin{equation}
{d a_i\over dt} = {2\over m_i} \sqrt{\frac{a_i}{GM_\star} } {\partial U
\over \partial \lambda_i},
\label{semar}\end{equation}
\begin{equation}
{d e_i\over dt} = - {1\over e_i m_i \sqrt{GM_\star a_i }} {\partial U
\over \partial \tilde{\omega}_i}.
\label{eccr} 
\end{equation}

\noindent The right hand side of equation~(\ref{eccr}) is written to
lowest order in the eccentricities which are assumed to be small.
Here, $U$ is the interaction potential which, to first order in the
eccentricities, is of the form:
\begin{equation}
U= -{m_1 m_2\over a_2}\left[ e_1 S_1 (\alpha)\cos\Phi_1 +e_2
S_2(\alpha)\cos\Phi_2 \right],
\label{rpot} \end{equation}
\noindent where $\alpha = a_1/a_2$ and the $S_i$ are known functions
(Goldreich~1965, Dermott et al. 1988), the detailed form of which does
not affect our conclusions.

The rates of change of $e_i$ and $a_i$ also have additional
contributions $(de_i/dt)_t$ and $(da_i/dt)_t$ arising from tidal
dissipation (hereafter, the subscript 't' will denote variations due
to tidal effects).  Thus, by use of equations~(\ref{semar})
and~(\ref{rpot}), we obtain:
\begin{equation}
{d a_1\over dt} = {2p\over m_1} \sqrt{\frac{a_1}{GM_\star}}F + \left({d
a_1\over dt}\right)_t,
\label{semar1}\end{equation}
\begin{equation}
{d a_2\over dt}  = -{2(p+1)\over m_2} \sqrt{\frac{a_2}{GM_\star} }F
+ \left({d a_2\over dt}\right)_t,
\label{semar2}\end{equation}
with:
\begin{equation}
F= {m_1 m_2\over a_2}\left[ e_1 S_1 (\alpha)\sin\Phi_1 +e_2
S_2(\alpha)\sin\Phi_2 \right].
\label{rpot1} \end{equation}

\noindent In addition, we find, from equations~(\ref{eccr})
and~(\ref{rpot}), for the rate of change of the eccentricities:
\begin{equation}
{d e_1\over dt} = - {1\over e_1 m_1 \sqrt{GM_\star a_1 }}F_1 +
\left({d e_1\over dt}\right)_t ,
\label{eccr1} \end{equation}
\begin{equation}
{d e_2\over dt} = - {1\over e_2 m_2\sqrt{GM_\star a_2 }}F_2 + \left({d
e_2\over dt}\right)_t ,
\label{eccr2} \end{equation}
where:
\begin{equation}
F_i= {m_1 m_2\over  a_2} e_i S_i (\alpha)\sin\Phi_i, \; \; \; \; (i=1,2)
\label{rpot2} \end{equation}
and we have $F= F_1 + F_2.$

We are now going to show from the above equations that, if the
resonance is maintained on average, the eccentricities must decrease
with time.  This means that the existence of the re\-so\-nance acting
through the two resonant angles cannot prevent the decay of the
eccentricities due to the action of the tides.  Indeed, for the
resonance to be maintained on average, we must have $d\ln (a_1/a_2)/dt
=0.$ Equations (\ref{semar1}) and (\ref{semar2}) then indicate that,
on average:
\begin{equation}
\left[{2p\over m_1a_1} \sqrt{\frac{a_1}{GM_\star}}+{2(p+1)\over m_2a_2}
\sqrt{\frac{a_2}{GM_\star} }\right] F = -\left[{d \ln(a_1/a_2) \over
dt}\right]_t.
\label{semarat}
\end{equation}

\noindent We can now use equations~(\ref{eccr1}) and~(\ref{eccr2}) to
obtain:
\begin{equation}
m_1 \sqrt{GM_\star a_1 }{d e_1^2\over dt}+ m_2 \sqrt{GM_\star a_2 }{d
e_2^2\over dt} = - 2F +m_1 \sqrt{GM_\star a_1 }\left({d e_1^2\over
dt}\right)_t +m_2 \sqrt{GM_\star a_2 }\left({d e_2^2\over
dt}\right)_t.
\label{eccrs} 
\end{equation}

\noindent Now, the effect of the tides causing circularization is to
decrease $e_i$, i.e. $(de_i/dt)_t<0$, while conserving the angular
momentum of $m_i$.  Thus $a_i$ also decreases.  But tides are stronger
in $m_1$, which is closer to the central star, which means that
$[d\ln(a_1/a_2)/dt]_t < 0$.  Equation~(\ref{semarat}) then implies
that $F$ is positive.  It then follows that the right hand side of
equation~(\ref{eccrs}) is negative from which we can deduce that the
eccentricities decrease with time as long as any one of them is non
zero.  This implies that if the resonance is maintained the $e_i$ must
ultimately decrease.

However, for the resonance to be maintained, one requires that the
change in $a_1$ produced by tides in one libration period be much
smaller than the amplitude of the oscillation in $a_1$ in resonance,
$\Delta a_1$ (adiabatic criterion, see, e.g., Dermott et al. 1988).
We are now going to show that this is not expected to be compatible
with a decrease of the eccentricities.  Indeed, when the eccentricity
decreases, the libration period increases and $\Delta a_1$ decreases,
so that at some point the resonance is broken. To see how this
happens, as long as the eccentricities are not too small, one can use
a perturbation scheme with small parameter
$\epsilon=\sqrt{m_i/M_\star}$, where $m_i$ is the largest of the
planet masses.  For resonance libration, we expect $d/dt = {\cal
O}(\epsilon n_i)$.  As $d \tilde{\omega}_i/dt = {\cal O}(\epsilon^2
n_i)$, to lowest order this is commonly neglected (Dermott et
al. 1988).  Under this scheme (which in addition requires $e_i \gg
\epsilon^{2/3}$), the angles $\Phi_1$ and $\Phi_2$ obey the same
equation (Dermott et al. 1988):
\begin{equation}
{d\Phi_1\over dt}= {d\Phi_2\over dt} = pn_1 - (p+1)n_2,
\label{rlib} \end{equation}
and  $\Phi_1 -\Phi_2 = \tilde{\omega}_1-\tilde {\omega}_2 $
is constant.  Thus we find:
\begin{equation}
{d^2\Phi_1\over dt^2}= -{3F\over \sqrt{GM_\star}}\left[ {p^2n_1\over m_1
 \sqrt{a_1}} +{(p+1)^2n_2\over m_2\sqrt{a_2}}\right]
 + p\left({dn_1\over dt}\right)_t -
(p+1)\left({dn_2\over dt}\right)_t .
\label{rlib1} \end{equation}

\noindent This can be further reduced to a forced pendulum equation
(e.g., Goldreich~1965):
\begin{equation}
{d^2\xi\over dt^2}= -\omega_L^2\sin\xi + p\left({dn_1\over dt}\right)_t -
(p+1)\left({dn_2\over dt}\right)_t ,
\label{rlib2} \end{equation}
where $\xi =\Phi_1-\delta,$ with:
\begin{equation} \tan \delta= {e_2S_2\sin(\tilde{\omega}_1-
\tilde {\omega}_2)\over
e_1S_1 + e_2 S_2\cos(\tilde{\omega}_1-\tilde {\omega}_2)},
\end{equation}
and the libration frequency is given by:
\begin{equation}\omega_L^2 = {3m_1 m_2\over \sqrt{GM_\star} a_2}
\left( {p^2n_1\over m_1 \sqrt{a_1}} +{(p+1)^2n_2\over
m_2\sqrt{a_2}}\right)\sqrt{e_1^2
S_1^2+e_2^2S_2^2+2e_1e_2S_1S_2\cos(\tilde{\omega}_1-\tilde
{\omega}_2)}.
\end{equation} 

\noindent From adiabatic invariance, we conclude that as the
eccentricities inevitably decrease because of circularization causing
a decrease in libration frequency, the oscillation amplitude of
$d\xi/dt$ decreases as do the excursions $\Delta a_i$ while the
amplitude of $\xi$ itself increases.  This naturally is expected to
lead to the disruption of the resonance.

We have shown that for the resonance to be maintained the
eccentricities have to decrease, and such a decrease of the
eccentricities is expected to lead to the resonance being disrupted.
Therefore, the resonance is not expected to be maintained under the
action of tidal circularization. However, secular evolution of
the semi--major axes occuring because of the circularization may cause
other resonances to be approached with subsequent repetition of the
disruption process discussed here.

Note that at the order we have worked above, $ \Delta \tilde{\omega} =
\tilde{\omega}_1-\tilde {\omega}_2$ is constant, corresponding to a
secular resonance for which apsidal alignment is maintained.  As this
type of resonance does not require a mean motion commensurability, it
is possible for it to be maintained during the circularization process
as well as some scattering and merger events (see below).

\section{Model and initial conditions}\label{sec:model}

We consider a system consisting of a primary star and $N$~cores or
protoplanets embedded in a gaseous disk surrounding it.  The cores
undergo gravitational interaction with each other and the star and are
acted on by tidal torques from the disk.
  
Work by several authors (e.g., Kley et al.~2004; Papaloizou~\&
Szuszkiewicz~2005; Cresswell \& Nelson 2006) has demonstrated that the
essential aspects of their motion can be captured by $N$--body
integration.  The effect of the disk torques and dissipative forces
are included in the integration.  Such a procedure has been shown to
give results very similar to those obtained when the disk response
induced by a planetary perturber and the torques acting back on
the protoplanet are calculated using hydrodynamic
simulations.

The equations of motion are:
\begin{equation} 
{d^2 {\bf r}_i\over dt^2} = -{GM_\star{\bf r}_i\over |{\bf r}_i|^3}
-\sum_{j=1\ne i}^N {Gm_j  \left({\bf r}_i-{\bf r}_j \right) \over |{\bf
    r}_i-{\bf r}_j |^3} -{\bf \Gamma} +{\bf \Gamma}_{i} +{\bf \Gamma}_{r} \; ,
\label{emot}
\end{equation} 

\noindent where $M_\star$, $M_i$ and ${\bf r}_i$ denote the mass of
the central star, that of planet~$i$ and the position vector of planet
$i$, respectively.  The acceleration of the coordinate system based on
the central star (indirect term) is:
\begin{equation} 
{\bf \Gamma}= \sum_{j=1}^N {Gm_j{\bf r}_{j} \over |{\bf r}_{j}|^3},
\label{indt}
\end{equation} 

\noindent and that due to tidal interaction with the disk and/or the
star is dealt with through the addition of extra forces as in
Papaloizou~\& Larwood~(2000):
\begin{equation}
{\bf \Gamma}_{i} = -\frac{1}{t_{m,i}} \frac{d {\bf r}_i}{dt} -
\frac{2}{|{\bf r}_i|^2 t_{e,i}} \left( \frac{d {\bf r}_i}{dt} \cdot
{\bf r}_i \right) {\bf r}_i - \frac{2}{ t_{i,i}}
\left( \frac{d {\bf r}_i}{dt} \cdot {\bf e}_z \right) {\bf e}_z,
\end{equation}

\noindent where $t_{m,i}$, $t_{e,i}$ and $t_{i,i}$ are the timescales
over which, respectively, the angular momentum, the eccentricity and
the inclination with respect to the unit normal ${\bf e}_z$ to the gas
disk midplane change.  Evolution of the angular momentum and
inclination is due to tidal interaction with the disk, whereas
evolution of the eccentricity occurs due to both tidal interaction
with the disk and the star.  We have:
\begin{equation}
\frac{1}{t_{e,i}} = \frac{1}{t_{e,i}^d} + \frac{1}{t_{e,i}^s} ,
\end{equation}

\noindent where $t_{e,i}^d$ and $t_{e,i}^s$ are the contribution from
the disk and tides raised by the star, respectively.  Relativistic effects
are included through ${\bf \Gamma}_{r}$ ( see Papaloizou \& Terquem 2001).

\subsection {Orbital circularization due to tides from the central star}
 
The circularization timescale due to tidal interaction with the star
is given by Goldreich~\& Soter (1966) as:
\begin{equation}
t_{e,i}^s = 4.065 \times 10^{4} \; \left( \frac{{\rm M}_\oplus}{M_i}
\right)^{2/3}  \left( \frac{20 a_i}{{\rm 1~au}} \right)^{6.5}  Q'
\; \; \; {\rm years} ,
\label{teccs}
\end{equation}

\noindent where $a_i$ is the semi--major axis of planet~$i.$ Here and
below we have adopted unit density in cgs~units for the planet. The
parameter $Q'= 3Q/(2k_2),$ where $Q$ is the tidal dissipation function
and $k_2$ is the Love number.  For solar system planets in the
terrestrial mass range, Goldreich \& Soter (1966) give estimates for
$Q$ in the range range 10--500 and $k_2 \sim 0.3.$ The values of these
quantities are clearly very uncertain under the very different
physical conditions likely to be appropriate to extrasolar planets.
We have performed simulations with $Q' = 10$,100~and~1000.  In the
first case, tidal circularization is very effective, while in the last
it only produces very small effects over practical simulation times.

 \noindent For a 1~earth mass planet at 0.05~au and $Q' = 100,$ we get
$t_{e,i}^s = 4 \times 10^6$ years, whereas at 0.1~au and for $Q' =
1000$, we get $t_{e,i}^s = 4 \times 10^9$ years.  Thus a range of
circularization timescales going from short compared to the formation
time to comparable to the age of the system may apply.

\subsection{Type~I migration}

In the local treatment of type~I migration (e.g., Tanaka et
al.~2002), if the planet is not in contact with the disk, there is no
interaction between them so that $t_{m,i}$, $t_{e,i}^d$ and $t_{i,i}$
are taken to be infinite.  When the planet is in contact with the
disk, disk--planet interactions occur leading to orbital migration as
well as eccentricity and inclination damping (e.g., Ward~1997).  In
that case, away from the disk edge, we adopt:
\begin{equation}
t_{m,i} = 146.0  \; \left[ 1+ \left( \frac{ e_i }{1.3 H/r}\right)^5 \right]
\left[ 1- \left( \frac{ e_i }{1.1 H/r}\right)^4 \right]^{-1} 
\; \left( \frac{H/r}{0.05}
\right)^2 \;
\frac{{\rm M}_\odot}{M_d} \;
\frac{{\rm M}_\oplus}{M_i} \; \frac{a_i}{{\rm 1~au}} \; \; \; {\rm years} ,
\label{tm}
\end{equation}
\begin{equation}
t_{e,i}^d = 0.362 \; \left[ 1+ 0.25 \left( \frac{ e_i }{H/r}\right)^3
\right] \; \left( \frac{H/r}{0.05}
\right)^4 \;
\frac{{\rm M}_\odot}{M_d} \; \frac{{\rm M}_\oplus}{M_i} \;
\frac{a_i}{{\rm 1~au}} \; \; \; {\rm years} ,
\label{te}
\end{equation}

\noindent and $t_{i,i}=t_{e,i}$ (eq.~[31] and~[32] of Papaloizou~\&
Larwood~2000 with $f_s=0.6$).  Here $e_i$ is the eccentricity of
planet~$i$, $H/r$ is the disk aspect ratio and $M_d$ if the disk mass
contained within 5~au.  We have assumed here that the disk surface
mass density varies like $r^{-3/2}$.  For a 1~earth mass planet on a
quasi--circular orbit at 1~au, we get $t_{m,i} \sim 10^5$~yr and
$t_{e,i} \sim 500$~yr for $M_d=10^{-3}$~M$_\odot$ and $H/r=0.05$.
Note that the timescales given by equations~(\ref{tm}) and~(\ref{te})
can be used not only for small values of $e_i$, but also for
eccentricities larger than $H/r$.  The eccentricity dependence of
these timescales is supported by the simulations of Cresswell~\&
Nelson~(2006).  Their absolute normalization can be varied by
scaling the disk surface density. We have checked that the general
simulation outcomes are robust to varying the ratio $t_{e,i}^d/t_{m,i}$
by a factor of three.

\subsection{Corotation torques}

\label{COROTO}
 
Type~I migration as discussed above is caused through the excitation
of density waves at Lindblad resonances.  However, torques due to
corotation resonances may also act. These depend on the gradient of
specific vorticity or vortensity (Goldreich \& Tremaine 1979).  They
are generally small, actually vanishing in a keplerian disk with a
surface density profile $\propto r^{-3/2},$ except where the surface
density varies fairly rapidly. Based on a three dimensional
linear response calculation, for a planet on a circular orbit and disk
surface density $\Sigma \propto r^{-\alpha,}$ Tanaka et al. (2002)
find that migration stops for $\alpha=-27/11$
%\begin{equation}
%\label{eq:tanaka}
%t_{m,i}  = (2.7+1.1\alpha)^{-1}\frac{M_{\odot}^2}{2\pi\Sigma a_i^2M_p}
%\left({H\over r}\right)^2P_{orb, i},
%\end{equation}
%where $P_{orb,i}$ is the orbital period of planet $i.$
%According to this, migration stops for $\alpha = - 27/11$ 
and is outward for surface density profiles that decrease more rapidly
inwards.

In our simulations, for simplicity, we have either adopted
equations~(\ref{tm}) and~(\ref{te}) with no interaction interior to
the disk inner edge, which from the above discussion we expect to
correspond to a moderate taper with $\alpha= -27/11$, or allowed for a
very sharp edge as described below.
 
Recently, Masset et al. (2006) have proposed that inward protoplanet
migration can be halted near sharp disk inner edges which act as
traps.  Here we study the migration of protoplanets into an inner
evacuated cavity so we thus consider the possibility of corotation
torques produced in a narrow region near the disk inner cavity
boundary.  We comment that the calculation of corotation torques is
very uncertain, being dependent not only on the details of the edge
profile, but also on the degree of resonance saturation which itself
depends on the amount of turbulence and viscosity present (Masset et
al. 2006). We here point out two possible effects that may act to
reduce the effectiveness of such edge torques when there is a system
of interacting planets. The first is planet--planet scattering, which
could move the semi--major axis of a planet across the edge. The
second is that protoplanets on eccentric orbits will only sample the
edge for a fraction of the orbit and accordingly suffer a reduced
torque.  Note that orbital eccentricity is more likely to be sustained
when the protoplanet orbit is only partly contained within the disk as
then the effectiveness of disk damping is reduced.

Corotation torques can act to produce outward torques in the inner
edge domain $R_{in}-\Delta_r/2 < r < R_{in}+\Delta_r/2 ,$ where the
surface density changes rapidly.  Here the edge is centered on
$r=R_{in}$ and the total width of the domain is $\Delta_r.$ As in the
case of Lindblad torques, the effect of orbital eccentricity is to
reduce such  corotation torques and this effect has been incorporated in our
modelling.  

Masset et al. (2006) indicate that when $a_i=R_{in},$ outward corotation
torques may exceed the normal inward type I torques by a factor of
five when $e=0.$ To investigate the possible role of such torques, in
some of our simulations we adopted the following approximate
procedure. In the edge domain, we replaced $t_{m,i}$ by
$-0.2t_{m,i,0}(1+e_ir/H),$ where $t_{m,i,0}$ denotes $t_{m,i}$
evaluated for $e_i =0.$ The factor in brackets accounts for the fact
that for large $e_i,$ the
effective values of the azimuthal number $m$ contributing to the
corotation torque are reduced by a factor $H/(re_i)$.  In practice,
this detail is not important for the simulations we carried out,
because $e_i$ never significantly exceeds $H/r$ in these cases.
Thus a planet in the center of the domain with $e_i =0$
experiences an outward torque of the required magnitude, while for
larger $e_i$ the time averaged torque will decline through the factor
$[H/(re_i)][\Delta_r/(2r e_i)].$ The second factor here estimates the
fractional reduction of the time spent in the edge domain once $e_i >
\Delta_r/(2R_{in}).$

Although we have focussed on the reversed edge torque as being due to
a corotation effect, it is possible that similar features could be
produced by, e.g., the presence of a toroidal magnetic field near the
inner edge (Terquem 2003).

\subsection{Numerical integration and initial conditions}

The equations of motion are integrated using the Bulirsch--Stoer
method (e.g., Press et al.~1993).  All the planets are supposed to
have an identical mass density $\rho=1$~g~cm$^{-3}$.
If the distance between planets~$i$ and~$j$ becomes less than
$[3M_i/(4 \pi \rho)]^{1/3} + [3M_j/(4 \pi \rho)]^{1/3}$, a collision
occurs and as is commonly assumed in studies of this kind,
the planets are assumed to merge.  They are subsequently
replaced by a single planet of mass $M_i+M_j$ which is given the
position and the velocity of the center of mass of planets~$i$
and~$j$.

The simulations begin by placing $N$ planets on coplanar orbits in an
annulus with outer and inner radii $r_{\rm out}$ and $r_{\rm
in}=xr_{out}$, respectively.  In some cases, the planets $i
=1,2,...,N$ were given the radial coordinate $r =
r_{in}[1+(x^{-3}-1)/(3i-2)]^{1/3}$ together with the polar angle $\varphi
=2\pi/i.$ In other cases, $r$ and $\varphi$ were chosen randomly.  The
planets were then given the local circular velocity in the azimuthal
direction.  We have fixed $r_{\rm out}=1$ or 2~au and $x=0.1$.  Note
that the range of radii over which the cores are initially spread does
not affect the outcome of the simulations.  Indeed, if $r_{\rm out}$
were larger, the cores would just take longer to migrate in.  The disk
is supposed to be truncated at some inner edge radius $R_{\rm in}$ in
the neighborhood of which a corotation torque may apply.

Initially, all the planets have the same mass $M_p$.  We have fixed
$M_p=0.1$~M$_\oplus$ or $M_p=1$~M$_\oplus$.  This range of masses has
been chosen because they have expected migration times from 5 au to
the central regions of the disk that are comparable to disk lifetimes.
Cores more massive than 1~M$_\oplus$ can subsequently form through
mergers.  We investigate what final systems of planets may be produced
interior to the disk inner boundary.

Time $t=0$ marks the beginning of the simulations.  It corresponds to
the time when the cores considered begin to migrate from the initial
positions allocated to them.  As it takes at least close to a million
years to form these cores, $t=0$ for the scenario envisaged here
should corrrespond to a time when the disk has already evolved
significantly.  Note that some cores may have begun to form further
away from the central star than the initial positions allocated to
them.  Here we assume that we can take $t=0$ to be the time at which
all cores formed in the disk that are able to migrate down to the
inner cavity within the disk lifetime are contained within the radius
$r_{\rm out}$ of the initial distribution. The total mass contained in
these cores is varied between 1 and 25 earth masses. We comment that
both increasing $r_{out}$ or decreasing the initial core mass have the
effect of extending the evolution time, that being inversely
proportional to the core mass.  Simulations performed with either
larger $r_{out}$ such as A2, A9, B3, or initial core masses reduced by
a factor of ten such as A10, indicated below, produce qualitatively
similar end results but with correspondingly reduced final total
masses in the latter case.  This is indicative that drawing out the
evolution time does not alter the qualitative behavior.
 
\section{Numerical results}
\label{sec:results}

For the runs presented here, we fixed $M_\star=1$~M$_\odot,$
$H/r=0.05$ and $M_d=10^{-3}$~M$_\odot.$ The prescriptions for disk
planet torques, eccentricity and inclination damping rates were, apart
from possible modifications listed in table~\ref{tab1}, as described
in section~\ref{sec:model}.  With these parameters specified, the
quantities characterizing a run were the initial number of planetary
cores, $N,$ their initial mass, $M_p,$ the tidal dissipation
parameter, $Q',$ the radius of the inner disk edge, $R_{\rm in},$ and
the bounding radii of the initial planet distribution, $r_{\rm in}$
and $ r_{\rm out}$.  Table~\ref{tab1} lists the parameters
corresponding to the different runs.  The initial number of planets
$N$ is either~10, 12 or~25.  The outer radius of the initial
distribution, $r_{out}$, is either~1 or~2~au.

\subsection{General outcome}

All the runs start with the same qualitative evolution.  A few
collisions and mergers take place very close to the beginning of the
simulation before significant migration occurs.  These result from the
initial unstable distribution of orbits.  Then as the planets migrate
inwards, further collisions and mergers occur on a timescale somewhat
shorter than the complete migration timescale.  Finally the runs end
with a stable configuration with a few planets, typically between 2
and 5, which will be inside the inner cavity when no edge corotation
torques are applied.  In that case, the most massive planets tend to
be on the tightest orbits, as they migrate faster.  Mean motion
resonances are always established during the migrating phase.  Some
rearrangement may take place as the planets approach the inner cavity
and further collisions occur, but by the time all the planets left
over finally enter the cavity, mean motion resonances between almost
all pairs of planets have been established.  At that stage, residual
scattering/mergers, should they occur, together with tidal interaction
with the central star leading to circularization of the orbits, on a
timescale which is shorter for closer in planets, results in the
disruption of strict commensurabilities (see
section~\ref{sec:disruption}).  The semi--major axes usually do not
evolve very significantly however, so the mean motions can stay near
commensurate.

\begin{figure}
%\centerline{
%\epsfig{file=figure1.ps,height=12cm,width=10cm }}
\includegraphics[height=12cm]{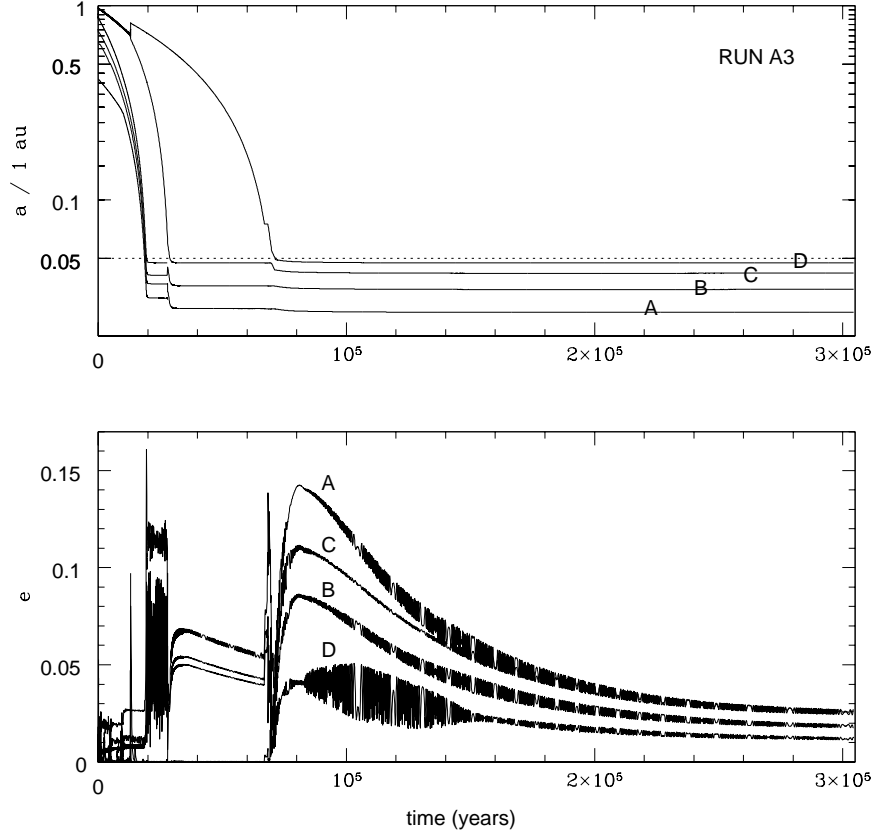}
\caption[]{ Evolution of the semi-major axes (in units of au and in
logarithmic scale; {\em upper plot}) and of the eccentricity ({\em
lower plot}) of the 12 planets in the system versus time (in units of
years) for run~A3. The solid lines correspond to the different
planets, each having an initial mass of 1~M$_\oplus$.  In this and
other similar figures, a line terminates just prior to a collision.
On the upper plot, the dotted line indicates the location of the inner
cavity ($R_{\rm in}=0.05$~au here).  The letters label the different
planets left after collisions have occurred.  The mass of planets~A,
B, C and D is 2, 6, 3 and 1~M$_\oplus$, respectively.  Planets~A
and~B, B and C, and C and D are in 3:2, 4:3, 6:5 mean motion
resonances, respectively.  }
\label{fig1}
\end{figure}

We now describe in more detail run~A3, which is a typical run.  It
begins with $N=12$~planets each having a mass $M_p=1$~M$_\oplus$ and
for which $r_{\rm out}=1$~au and $R_{\rm in}=0.05$~au.  The time
evolution of the semi--major axes and eccentricities of the planets up
to $3 \times 10^5$~years is shown in figure~\ref{fig1}.  Within the
first 100~years after the start of the calculation, before any
migration has occurred, 5~pairs of planets merge.  Another merger
occurs after $\sim 10^4$~years.  After $\sim 2 \times 10^4$~years, the
4~innermost planets enter the inner cavity, where their semi--major
axes do not evolve anymore.  When one of the planets still in the
disk finally approaches the inner cavity, after $\sim 3 \times
10^4$~years, it pumps up the eccentricity of the innermost planets
which results in 2~pairs of planets undergoing collisions and mergers.
At that point, 3~planets are left in the cavity, where they are joined
after $\sim 7 \times 10^4$~years by the outermost planet.  Pairs of
orbits are then in mean motion resonances.  Subsequent tidal
circularization, that acts on a timescale of a few millions years and
is seen on the bottom panel of figure~\ref{fig1}, will disrupt the
resonances, but the orbits may stay nearly commensurate. 

To study the action of tidal circularization on a commensurability
formed by disk planet interaction, for purely illustrative purposes we
consider a simple exam\-ple with two planets.  In the notation of
section~\ref{sec:disruption}, these had masses $m_1= 2$~M$_{\oplus}$
and $m_2 = 8$~M$_{\oplus}.$ We performed two simulations with the same
disk parameters as A3 but, for practical reasons, the circularization
rates were taken to be ten and a hundred times faster.  Comparison of
these cases indicates that the form of the evolution is the same but
with the time scale appropriately stretched. These planets initially
migrated into the inner cavity and formed a 5:4 commensurability for
which the resonant angle $\Phi \equiv -\Phi_1 =
5\lambda_2-4\lambda_1-\varpi_{1}$ has a small libration about
zero. The subsequent evolution under the action of orbital
circularization is shown in figure~\ref{fig2a}.  As expected, the
libration amplitude increases as the planets begin to move out of the
5:4 resonance. In the case with faster circularization, the angle
$\Phi$ eventually begins to show circulation with the 4:3 resonance
being approached. In this particular example, the eccentricity attains
very small values $\le 0.001$ while away from the center of
resonances. The indication is that the system separates as it moves
away from resonant configurations associated with high
eccentricities.

\begin{figure}
%\centerline{
%\epsfig{file=figure2.ps,height=8cm,width=10cm}}
\includegraphics[height=12cm]{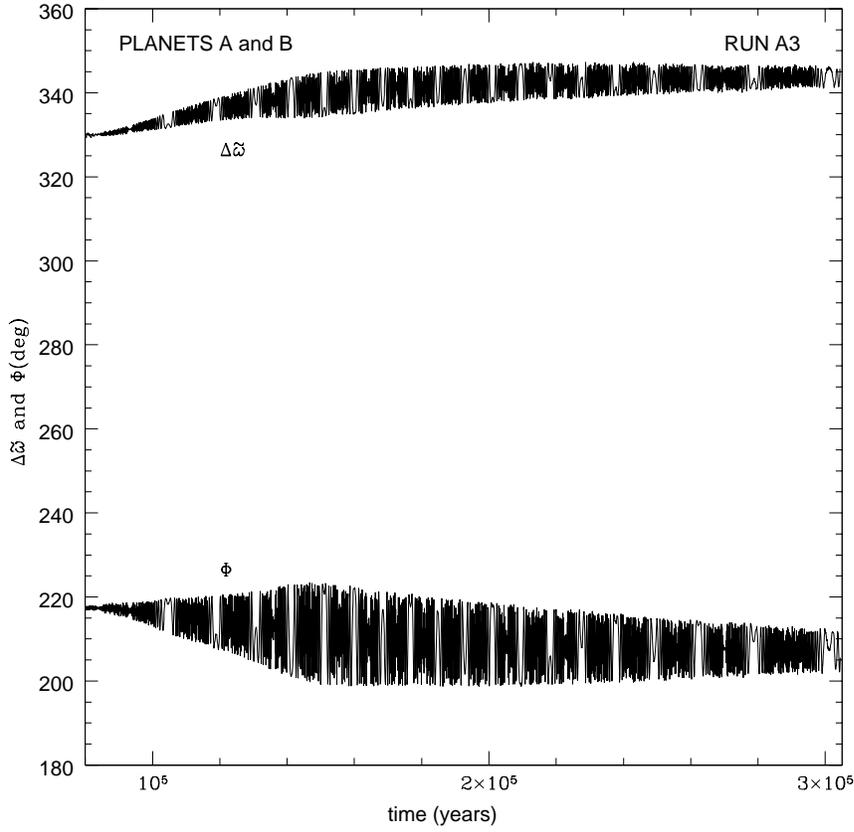}
\caption[]{ Evolution of the angular difference of the apsidal lines
$\Delta \tilde \omega$ and of the resonant angle $\Phi$ (in degrees)
for planets~A and~B versus time (in years), starting at $8 \times
10^4$~years after the beginning of the simulation, for the same run as
in figure~\ref{fig1}.  The angles librate about some fixed values
with an amplitude of a few degrees, which indicates mean motion
resonance (3:2 here) and apsidal locking.}
\label{fig2}
\end{figure}

\begin{figure}
%\centerline{
%\epsfig{file=figure2.ps,height=8cm,width=10cm}}
\includegraphics[width = 17cm, height=17cm, angle=270]{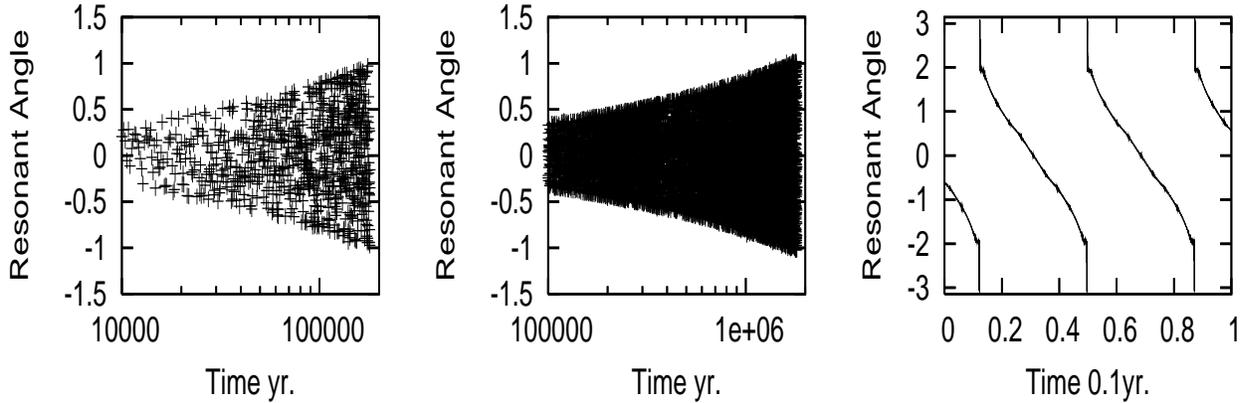}
\caption[]{ Evolution of the resonant angle $\Phi \equiv -\Phi_1 =
5\lambda_2-4\lambda_1-\varpi_{1}$ expressed in radians for two
interacting planets that disk interaction caused to enter a 5:4
resonance under the action of orbital circularization.  The left panel
is for a case with circularization rate $10$ times faster than that
illustrated in the middle panel. Both panels show the early stages of
the evolution during which the libration amplitude increases as the
planets begin to move out of the 5:4 resonance. Note that the
evolution illustrated in the middle panel is ten times slower than
that ilustrated in the left panel, showing that the evolution is driven
by the orbital circularization. In the case with faster
circularization, the angle $\Phi$ begins to show circulation after a
time $\sim 7\times 10^5$~yr, as the system moves towards the 4:3
resonance.  The right hand panel shows this circulation in a high time
resolution plot taken after $\sim7.4\times 10^5$~yr.  The circulation
period $\sim 12$~days, while the orbital period of the inner planet $\sim
3$~days. }
\label{fig2a}
\end{figure}

 For  run~A3, the 4~planets left at the end of the run, that we
label 'A', 'B', 'C' and 'D', have a mass of 2, 6, 3 and 1~M$_\oplus$,
respectively.  The two innermost planets, A and B, are in a 3:2 mean
motion resonance, with $|n_A/n_B-3/2| \sim 10^{-3}$, where $n_A$ and
$n_B$ are the mean motions of planets~A and~B, respectively.  In
figure~\ref{fig2} we plot the angular difference of the apsidal lines
$\Delta \tilde{\omega}$ and the resonant angle $\Phi=3 \lambda_B - 2
\lambda_A - \tilde{\omega}_A$, where $\lambda_A$ and $\lambda_B$ are
the mean longitudes of planets~A and~B, respectively, and
$\tilde{\omega}_A$ is the argument of pericentre of planet~A.  After
$\sim 10^5$~years, both $\Delta \tilde{\omega}$ and $\Phi$ librate
about some fixed values (344 and 207~$\deg$, respectively) with an
amplitude of a few degrees, which indicates apsidal locking and mean
motion resonance.  Note that $\Delta \tilde{\omega}$ and $\Phi$ do not
necessarily librate about 0 or 180~$\deg$ when there is a mean motion
resonance.  As shown by Beaug\'e et al. (2003), the equilibrium value
of these angles tend to depart from either 0 or 180~$\deg$ when the
eccentricity of the planets is not small (typically higher than $\sim
0.1$).

We have run a case with a lower total mass (A10, $N=10$ and
$M_p$=0.1~M$_\oplus$) and a case with a higher total mass (A11,
$N=25$ and $M_p$=1~M$_\oplus$).  The results of these runs are similar
to those described above, only the mass of the planets left in the
inner cavity changes, roughly scaling with the initial total mass. 

In the run A3 described above, we found that $\Delta \tilde \omega$
and $\Phi$ were librating about some fixed values.  Note that this is
not always the case.  In some of the runs we have performed, these
angles circulate, so that the systems are not formally exactly
commensurable.  We also point out that the commensurabilities given in
table~\ref{tab1} are accurate to within at least 1\%, and often to
within 0.1\%.

\subsection{Tidal circularization}

To investigate the effect of changing the orbital circularization
rate, we performed simulations A8 with $Q' =10$ and B2 with $Q' =
1000.$ Both these runs have inner cavity radius $R_{\rm in}=0.1$~au.
In the former case, orbital circularization is manifest in the
simulation, while in the latter case, it is too long to be manifest.

\begin{figure}
%\centerline{
%\epsfig{file= W3plot.ps,height=12cm,width=10cm,angle=270}}
\includegraphics[width=11cm,angle=270]{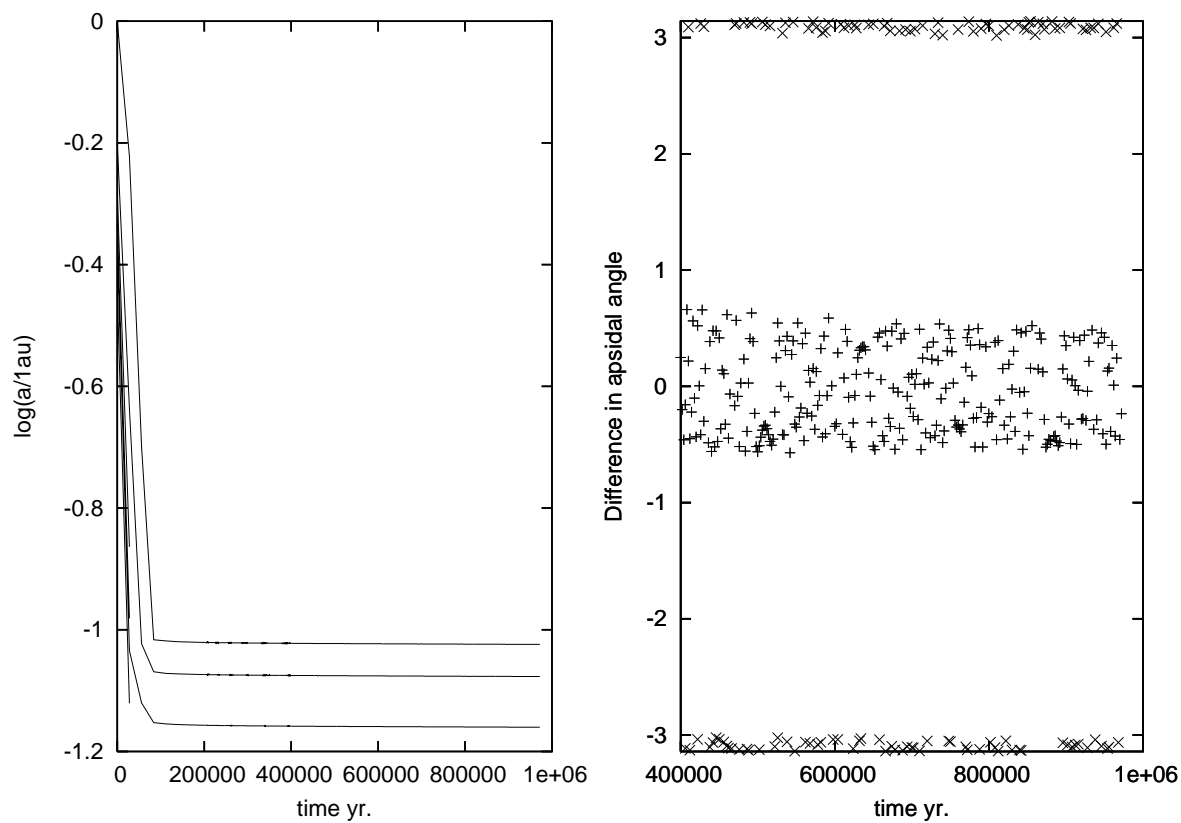}
\caption[]{The left panel shows, as in figure~\ref{fig1}, the
evolution of the semi--major axes for run A8. The right panel shows
the evolution of the angular differences of the apsidal lines $\Delta
\tilde \omega$ (in radians) for the two innermost, according to
semi--major axis, planets (points clustered about 0) and the innermost
and outermost of the three planets (points clustered about $\pi$ and
$-\pi.$ Note that because of periodicity any multiple of $2\pi$ can be
added.}
\label{fig4}
\end{figure}

\begin{figure}
%\centerline{
%\epsfig{file= W3C.ps,height=12cm,width=10cm,angle=270}}
\includegraphics[width=11cm,angle=270]{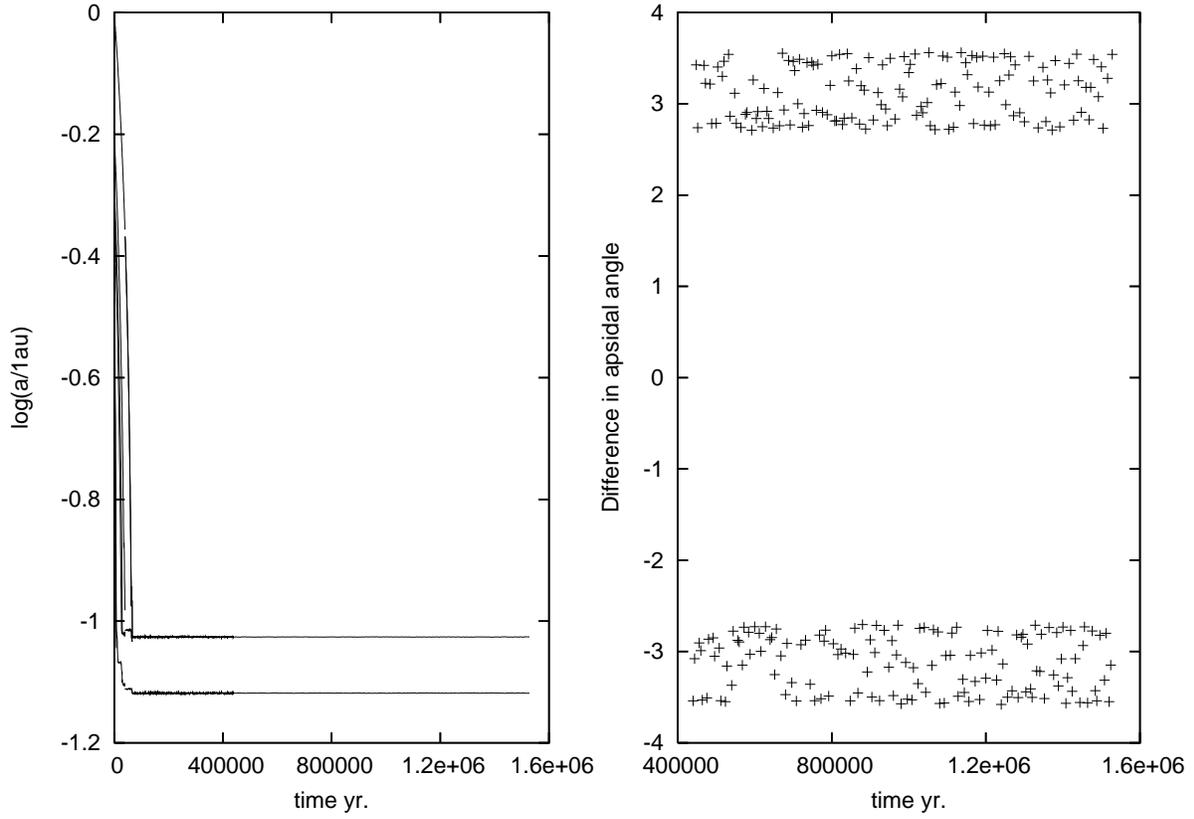}
\caption[]{The left panel shows, as in figure~\ref{fig4}, the
evolution of the semi--major axes for run B2 . The right panel shows
the evolution of the angular difference of the apsidal lines $\Delta
\tilde \omega$ (in radians) for the remaining two planets. This
oscillates around the anti--aligned position.}
\label{fig5}
\end{figure}

\begin{figure}
%\centerline{
%\epsfig{file=EccW3W3C.ps,height=12cm,width=10cm,angle=270}}
\includegraphics[width=11cm,angle=270]{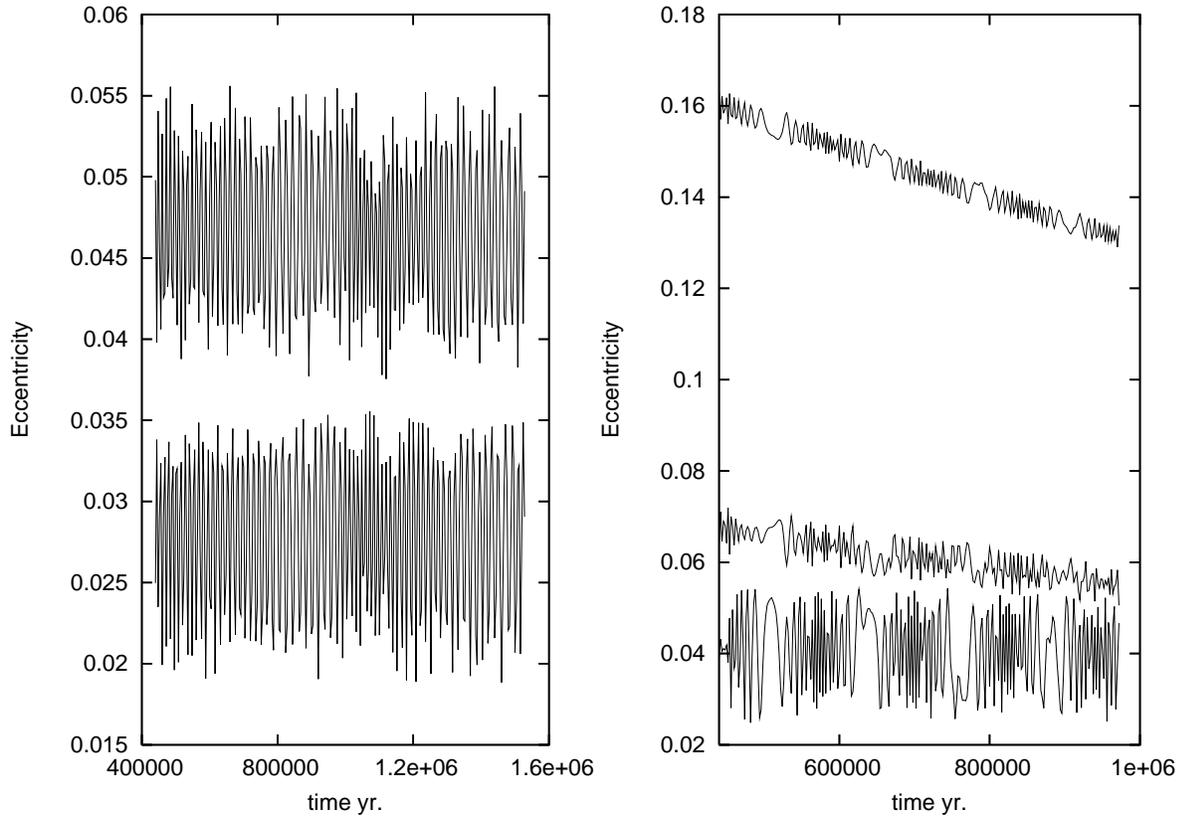}
\caption[]{The evolution of the eccentricities of the planets in the
runs B2 ({\em left panel}) and A8 ({\em right panel}).  In the former case,
the upper curve corresponds to the innermost planet according to
semi--major axis.  In the latter case, the lowest curve corresponds to
the outermost planet and the middle curve to the innermost planet. }
\label{fig6}
\end{figure}

The simulation A8 ended with three planets in the inner cavity in near
but not exact commensurability.  The evolution of the semi--major axes
is shown in figure~\ref{fig4} as is the evolution of the angular
differences of the apsidal lines $\Delta \tilde \omega$ for the two
innermost planets and the innermost and outermost of the three
planets. These librate about alignment in the former case and
anti--alignment in the latter case.  In figure~\ref{fig5}, the
evolution of the semi--major axes is shown for run B2 . The right
panel of this figure shows the evolution of the angular difference of
the apsidal lines for the two planets which remain in the inner
cavity.  This oscillates around the anti--aligned position. This case
has a very long circularization time.  It is possible that for some of
these cases with larger cavity radii, some orbital eccentricity
remains on $10^9$~years timescales, in which case the
alignment/anti--alignment of the apsidal lines could be observed.  The
evolution of the eccentricities in the simulations B2 and A8 is shown
in figure~\ref{fig6}. In the latter case, a small amount decay for the
innermost planets can be seen, while in the former, the circularization
rate due to tides induced by the central star is too small to have any
effect.

We also performed simulation B1 which had identical parameters to A8
but inner cavity radius $R_{\rm in}=0.05$~au and simulation B3 which
was identical to B2 apart from the size of the initial domain in which
the planets were started. In both simulations, the number of remaining
planets  and final period ratios were similar.

\subsection{Migration halted at the disk inner edge by corotation torques}

In the runs presented above, the interaction between the planets and
the disk leads to inward migration of the planets.  After a planet
enters the inner cavity, it is no longer pushed in by the disk, but it
can still be pushed in by planets further away which enter the cavity
at a subsequent time.  Note that when a planet approaches the inner
cavity, it may gently push in the planets which are already inside,
 but it may also perturb them in such a way that a merger
occurs.  Both processes are seen in the run~A3 displayed in
figure~\ref{fig1}.  

It has been argued recently (Masset et al.~2006) that when an embedded
planet reaches a region of the disk where the mass density decreases
sharply, the tidal torque from the disk is reversed so that migration
is halted and the planet is trapped at this location.  We have
performed runs in which the torque at the inner edge of the disk is
reversed according to the prescription described in section
\ref{COROTO} to test whether planets penetrate inside the cavity. In
simulations C1, C2 and C3, such  torques were applied.  The
boundary torques that we applied were strong enough to prevent entry
into the inner cavity while the disk was present.  We emphasize that
this is the important feature of the torque prescription
that we adopted and that otherwise results should be independent of
details.  

We considered two phases in these cases for which the disk edge torque
prevented entry into the inner cavity. While embedded in the disk, the
strong orbital circularization allowed close commensurabilities to
form among six to seven planets in a similar manner to that described
for simulation A3.  These are indicated in table~\ref{tab1}. A state
was reached for which the semi--major axes became almost constant such
that angular momentum transferred from inner to outer planets
prevented their inward migration.  After this state was reached, the
disk was removed (by setting the induced migration and circularization
rates to zero) in order to study the further evolution and in
particular the effect of removing the stabilizing influence of disk
eccentricity damping.  

\begin{figure}
%\centerline{
%\epsfig{file=CORPLOT.ps,height=12cm,width=10cm,angle=270}}
\includegraphics[width=11cm,angle=270]{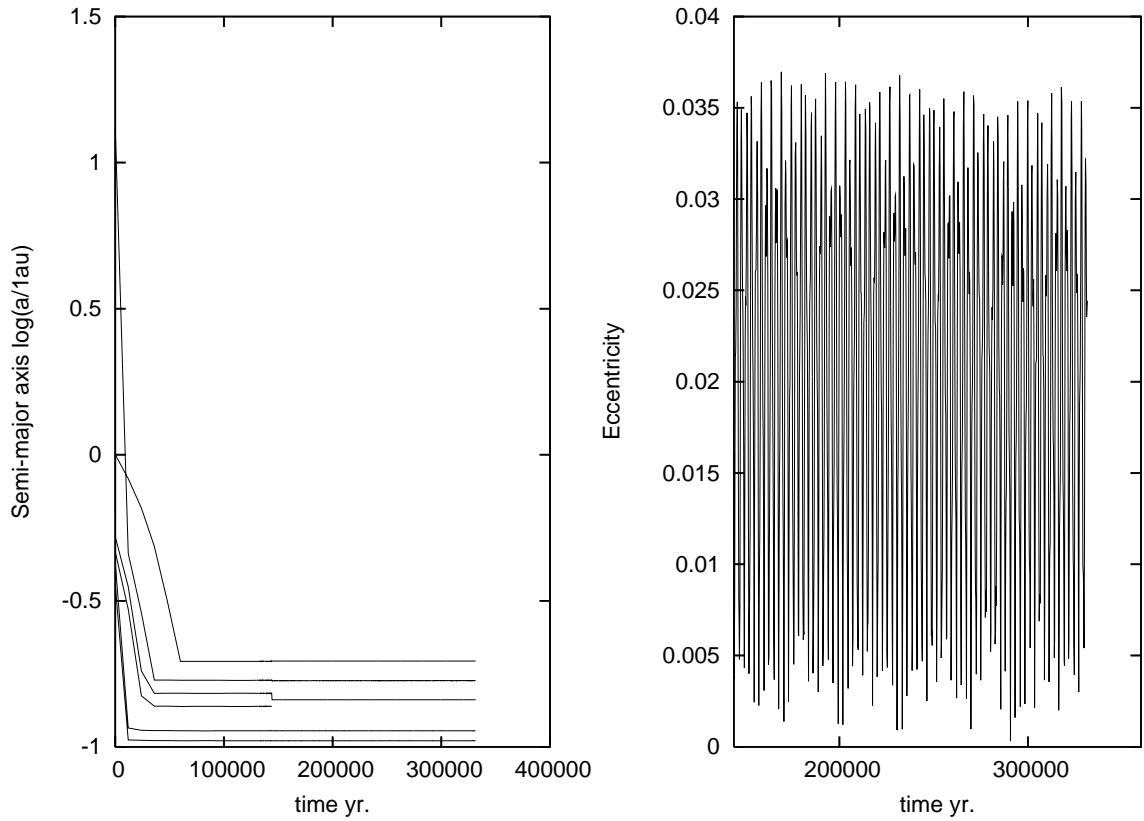}
\caption[]{The evolution of the semi--major axes ({\em left panel})
and the later evolution of the eccentricity of the outermost planet
({\em right panel}) for run C3. In this case, the disk was removed at
time $1.32 \times 10^5$~yr.  Shortly after that time, two of the
planets merged leaving a system of five planets.  }
\label{fig7}
\end{figure}

Simulation C3 had $Q' = 1000$ and the larger inner cavity radius
$R_{\rm in} =0.1$~au.  The evolution of the semi--major axes and
eccentricity of the outermost planet are shown in figure~\ref{fig7}.
As indicated in table~\ref{tab1}, this run produced a system of six
planets stably locked in a series of commensurabilities (9:8, 4:3,
7:6, 7:6, 5:4). In this state, the negative torques acting on the
outer planets were effectively balanced by the corotation torque
acting on the innermost planet so that evolution of the semi--major
axes ceased.  The disk was removed at time $1.32 \times 10^5$~years.
Shortly after that time, two of the planets merged leaving a system of
somewhat more widely spaced five planets that survived with almost
constant semi--major axes (see figure~\ref{fig7}). The period ratios
moving outwards were then (1.125, 1.44, 1.25, 1.26). The stability of
systems of low mass planets has been considered by Chambers et
al. (1996).  They considered systems of three objects which in our
case would correspond to 3~M$_{\oplus}$ planets for up to $\sim 10^7$
inner orbits, and we have evolved our systems for similar or longer
times. They found that the systems must be more widely separated to
ensure stability for longer times.  This is of course a statistical
statement, there are no guarantees in specific cases. We also note the
additional potential stabilisation provided by orbital circularization
in our case.  Nonetheless, if one makes the very arbitrary assumption
that their results can be simply extrapolated to $\sim 10^{11}$ inner
orbits, which would correspond to Gyr time scales, period ratios of
$\sim$~1.25 would be required. This corresponds to a spacing between
planets of $10.75$ Hill radii.  This is similar to what our systems
show except that, in the case of run~C3, the innermost pair are very
close to a 9:8 commensurability.  Although these two planets
maintained locked apsides, the appropriate resonant angle $\Phi$
showed long term variations but did not librate.  It is possible that
some of the planets in such systems could later merge, forming a more
widely separated system with fewer planets, but the general character
is likely to be preserved.  Simulations C1 and C2, which had smaller
inner cavity radii, led to similar configurations while the disk was
present. However, in these cases, the systems remained stable when the
disk was removed. This may be because of the increased importance of
orbital circulation, especially in the case of C2 which had seven
remaining planets.  In this case, with $Q'=10,$ orbital
circularization cannot be neglected in the simulation run time and
might be expected to assist system stability by preventing the slow
build up of orbital eccentricities that could result in orbit
crossing.

\section{Summary and Discussion}
\label{sec:disc}

We have calculated the evolution of a population of cores/planets with
masses in the range 0.1--1~M$_\oplus$ embedded in a disk.  They evolve
due to gravitational interaction with the central star, mutual
gravitational interactions, tidal interaction with the disk and the
star.  Mutual interactions lead to orbit crossing and mergers, so that
the cores grow during their evolution.  Interaction with the disk
leads to orbital migration.  As cores with different masses and at
different locations migrate with different rates, they capture each
other in mean motion resonances. Such captures enable planets to
migrate inside the cavity interior to the disk inner edge.  As they
approach closer to the central star, for small enough cavities their
orbits are circularized through tidal dissipation and strict
commensurabilities are lost. That process may be also aided by
scatterings and mergers of planets on unstable orbits that occur
interior to the disk inner edge.  Near--commensurability however may
be maintained.  Note too that if apsidal locking is established during
migration, it can be preserved through the operation of these
processes.  All the simulations end with a population of typically
between two and five planets, with masses depending on the initial
mass.  Note that the disk tidal torque may be reversed near the disk
inner edge.  When this is the case,  although it is possible 
that some planets can
still penetrate inside the cavity due to scatterings and/or weakening
of the torques because of a finite eccentricity,  some planets 
may be left in the disk, just beyond the inner edge, until that
disperses.

The qualitative results do not depend on the detail of the initial
conditions.  As long as a population of cores is able to migrate
inwards at different rates, the system evolves toward a family of a
few planets which are almost always on near--commensurate orbits.

The orbital migration and eccentricity damping timescales we have
adopted in this paper have been derived for type~I migration in
inviscid disks.  Note that type~I migration has been shown to follow a
random walk in a turbulent disk (Nelson \& Papaloizou 2004).  The
studies done in the present paper assume that the cores can migrate
down to the disk inner edge, in regions where the gas is ionized and
magnetic turbulence can develop (Fromang et al. 2002).  These studies
would of course not apply if there were no systematic inwards
migration of the cores/protoplanets of the type we consider here.  As
of today, there is no indication that type~I migration in a turbulent
disk has a systematic trend (Nelson 2005), but this cannot be ruled
out either.  It is likely that type~I migration does depend at least
to some extent on the torque exerted by the material that corotates
with the planet, which so far has not been taken into account in the
simulations, which lack the required resolution.  If there is a
systematic trend, as long as the time averaged migration rate is
inwards, by the averaging principle and as confirmed in test studies,
short term fluctuations do not qualitatively change the results we
have presented in this paper.

It has very recently been suggested (Paardekooper \& Mellema 2006)
that, because of effects arising from radiation trapping, type~I
migration in an optically thick laminar disk could be outward.  This
result, should it be confirmed for both laminar and turbulent disks,
suggests that cores would not migrate inwards in the disk inner parts
as long as the dust opacity is high enough there.  However, after the
dust settles and agglomerates to form cores, the opacity decreases and
migration could then resume.  It is indeed an observational fact that
the disk inner parts (whithin a few au) become optically thin before
the rest of the disk is depleted.

The calculations done in this paper show that if hot super--Earths or
Neptunes form by mergers of inwardly migrating cores, then such
planets are most likely not isolated.  We would expect to always find
at least one, more likely a few, companions on close and often
near--resonant orbits.  To test this hypothesis, it would be of
interest to look for planets of a few to $\sim 10$~earth masses in
systems where hot super--Earths or Neptunes have already been found
and there is no destabilizing influence of a giant planet close by.

It has been speculated that the cores of giant planets could form in a
way similar to that investigated here, by accumulation of cores in the
disk inner parts (e.g., Papaloizou \& Terquem 1999).  The calculations
presented in this paper suggest that to assemble a massive core in the
inner disk, significantly more mass in smaller cores may be needed to
begin with.  Indeed, most of the simulations end with at least three
planets, and not with a single planet containing all the initial mass.

\newpage 

\begin{table}
%\begin{center}
\begin{small}
\begin{tabular}{llllcclll} \hline \hline
Run & $N$ & $M_p$  & $Q'$ &  
$R_{\rm in }$  & $r_{\rm in}$  & $ r_{\rm out}$ &
Final masses  &  Final period ratios
\\
\hline 
  A1 & 10  & 1  &100 &0.05&  0.1 & 1 
&  4, 2, 3, 1 & 3:2,  4:3,  6:5  \\
  A2 & 10  & 1  &100 &0.05&  0.2 & 2 &  3, 6, 1    
& 4:3, 7:5  \\
  A3 & 12  & 1  &100 &0.05&  0.1 & 1 &  2, 6, 3, 1 
& 3:2,  4:3,  6:5   \\
  A4 & 12  & 1  &100 &0.05&  0.1 & 1 &  2, 6, 4    
& 4:3,  3:2  \\
  A5 & 10  & 1  &10 &0.05&  0.1 & 1 & 9 ,1       
&  2:1  \\
  A6 & 10  & 1  &1000 &0.05&  0.1 & 1 & 8 ,1 , 1    
&  1.46, 1.69  \\
  A7 & 10  & 1  &1000 &0.1&  0.1 & 1 &  8, 2   
&  1.42 \\
  A8 & 10  & 1  &10 &0.1&  0.1 & 1 & 6 ,3  ,1  
& 4:3, 6:5  \\
  A9 & 10  & 1  &1000 &0.1&  0.2 & 2 &  5 ,5    
&  1.86  \\ 
  A10 & 10  & 0.1  &100 &0.05&  0.1 & 1 &  0.9 ,0.1 
&  7:5 \\
  A11 & 25 & 1 & 100 & 0.05 & 0.1 & 1 & 12, 9, 4 
&  2:1, 11:8 \\
  B1& 10  & 1  &10 &0.05&  0.1 & 1 & 3 ,6 ,1     
&  1.37,  5:4 \\
  B2 & 10  & 1  &1000 &0.1&  0.1 & 1 & 4 ,6  &  1.37 \\
  B3 & 10  & 1  &1000 &0.1&  0.2 & 2 &  3 ,7      
&  1.37   \\
  C1& 10  & 1  &1000 &0.05&  0.1 & 1 &  (3, 3, 1, 1, 1, 1)       
& (7:6, 4:3, 6:5, 5:4, 5:4)  \\
     &     &    &     &    &     &   &    3, 3, 1, 1, 1, 1 
&  7:6, 4:3, 6:5, 5:4, 5:4   \\ 
  C2 & 10  & 1  &10 & 0.05&  0.1 & 1 &  (3, 2, 1, 1, 1, 1, 1) 
&  (6:5, 6:5, 5:4, 6:5, 6:5, 5:4)\\
    &     &    &     &    &     &   &    3, 2, 1, 1, 1, 1, 1
&  6:5, 6:5, 5:4, 6:5, 6:5, 5:4\\ 
  C3 & 10  & 1  &1000 & 0.1&  0.1 & 1 & (3, 3, 1, 1, 1, 1)     
&(9:8, 4:3, 7:6, 7:6, 5:4)  \\
     &     &    &     &    &     &   &    3, 3, 2, 1, 1   
&       9:8, 1.44, 5:4, 1.26    \\ 
\hline
\hline
\end{tabular}
\end{small}
%\end{center}
\caption{\small \label{tab1} This table lists the parameters for each
simulation.  The label~A denotes standard runs as described in
section~\ref{sec:model}, the label~B indicates that the disk
eccentricity damping rates were decreased by a factor of three, the
label~C denotes that edge corotation torques were applied.  $N$ is the
initial number of planets, $M_p$ is their mass in units of M$_\oplus$,
$Q'$ is the tidal dissipation parameter, $R_{in}$ is the radius of the
disk inner edge (in au), $r_{\rm in}$ and $ r_{\rm out}$ are the
bounding radii of the initial planet distribution (in au).  The table
also lists the final masses (in units of M$_\oplus$) of the objects
left at the end of the run from the innermost to the outermost planet
and the period ratios for neighboring final objects starting from the
innermost.  In the case of runs labelled~C, masses and period ratios
at the stage when the planets reached a steady configuration while
embedded in the disk are indicated in brackets.  Values subsequently
attained after disk removal are given on the lines below.  Indicated
mean motion commensurabilities apply to within one percent.}
\end{table}

\newpage

\end{document}